\documentclass[times, 10pt,twocolumn]{article} 
\usepackage{latex8}
\usepackage{times}

\usepackage{cite}
\usepackage[pagebackref=true,breaklinks=true,colorlinks,bookmarks=false]{hyperref} %hyperlinks
\usepackage{url}            
\usepackage{float}
\usepackage{graphicx}
\usepackage{subfig}
\usepackage{amsmath}
\usepackage{array}
\usepackage{multirow}
\usepackage{flushend}
\usepackage{amssymb}
\usepackage{amsfonts}
\usepackage{ulem}
\usepackage{booktabs}
\usepackage{textcomp}

\begin{document}

\title{Sparse Tensor-based Point Cloud Attribute Compression}

\author{Jianqiang Wang, Zhan Ma\\
Nanjing University, Nanjing, China\\
wangjq@smail.nju.edu.cn, mazhan@nju.edu.cn\\
% For a paper whose authors are all at the same institution, 
% omit the following lines up until the closing ``}''.
% Additional authors and addresses can be added with ``\and'', 
% just like the second author.
% \and
% Second Author\\
% Institution2\\
% First line of institution2 address\\ Second line of institution2 address\\ 
% SecondAuthor@institution2.com\\
}

\maketitle

\begin{abstract}
Recently, numerous learning-based compression methods have been developed with outstanding performance for the coding of the geometry information of point clouds. 
On the contrary, limited explorations have been devoted to point cloud attribute compression (PCAC). Thus, this study focuses on the PCAC by applying sparse convolution because of its superior efficiency for representing the geometry of unorganized points.
The proposed method simply stacks sparse convolutions to construct the variational autoencoder (VAE) framework to  compress the color attributes of a given point cloud.
To better encode latent elements at the bottleneck, we apply the adaptive entropy model with the joint utilization of hyper prior and autoregressive neighbors to accurately estimate the bit rate.  The qualitative measurement of the proposed method already rivals  the latest G-PCC (or TMC13) version 14 at a similar bit rate. And,  our method shows clear quantitative improvements to G-PCC version 6, and largely outperforms existing learning-based methods, which promises encouraging potentials for learnt PCAC.
\end{abstract}

\Section{Introduction}
\label{sec:intro}

Point cloud is a popular media format to represent 3D objects and scenes. It consists of a collection of points in a  3D space, and each point has its coordinate and associated attributes such as color, reflectance, etc, by which it can realistically and effectively represent arbitrary 3D objects/scenes. Thus the point cloud is used in vast applications like autonomous navigation, virtual reality, etc, which urges the development of high-efficiency Point Cloud Compression (PCC) for better service enabling.

Usually, point cloud geometry and attribute (e.g., color) are separately processed.
The former one focuses on the compression of point coordinates, and the latter one is for the coding of associated attributes. Normally, geometry compression is independently executed first, and then attribute compression is often conditioned on decoded geometry information. This paper tackles the color attribute compression, and assumes losslessly coded geometry for simplicity. Note that the compression of geometry components can be fulfilled using either standardized G-PCC~\cite{gpcc2021description} or learnt methods~\cite{Wang2020MultiscalePC, Wang2021SparseTM}.

{\bf Existing PCAC Methods.} Popular PCAC approaches mostly rely on rules-based transformation or prediction for efficient and compact color attribute representation such as the Region-Adaptive Hierarchical Transform (RAHT)~\cite{Queiroz2016CompressionO3},  Graph Fourier Transform (GFT)~\cite{Zhang2014PointCA},  hierarchical nearest-neighbor prediction based Lifting Transform~\cite{gpcc2021description}, etc. These methods have demonstrated convincing efficiency and some of them have been adopted in MPEG G-PCC standard~\cite{gpcc2021description,graziosi2020overview}.

On the other hand, following the successful application of deep neural networks (DNN) based image and video compression~\cite{balle2018variational, minnen2018joint, chen2021NLAIC}, learnt PCC has been extensively studied in both academia and industry. Particularly, both MPEG and JPEG in ISO/IEC task forces have been actively pursuing along this direction. Whereas, most learnt PCC methods focus on the compression of point cloud geometry occupancy using different 3D representation models, such as the 3D voxel grid~\cite{Quach2020ImprovedDP, Guarda2021AdaptiveDL, Nguyen2021LosslessCO}, octree model~\cite{Que2021VoxelContextNetAO, Huang2020OctSqueezeOE}, sparse tensor~\cite{Wang2020MultiscalePC, Wang2021SparseTM}, etc. Because of the powerful representative capacity of DNNs, these learnt point cloud geometry compression (PCGC) methods have shown even better coding performance than traditional rules-based methods like G-PCC.

\begin{figure*}[t]
	\begin{center}
	\includegraphics[width=6.2in]{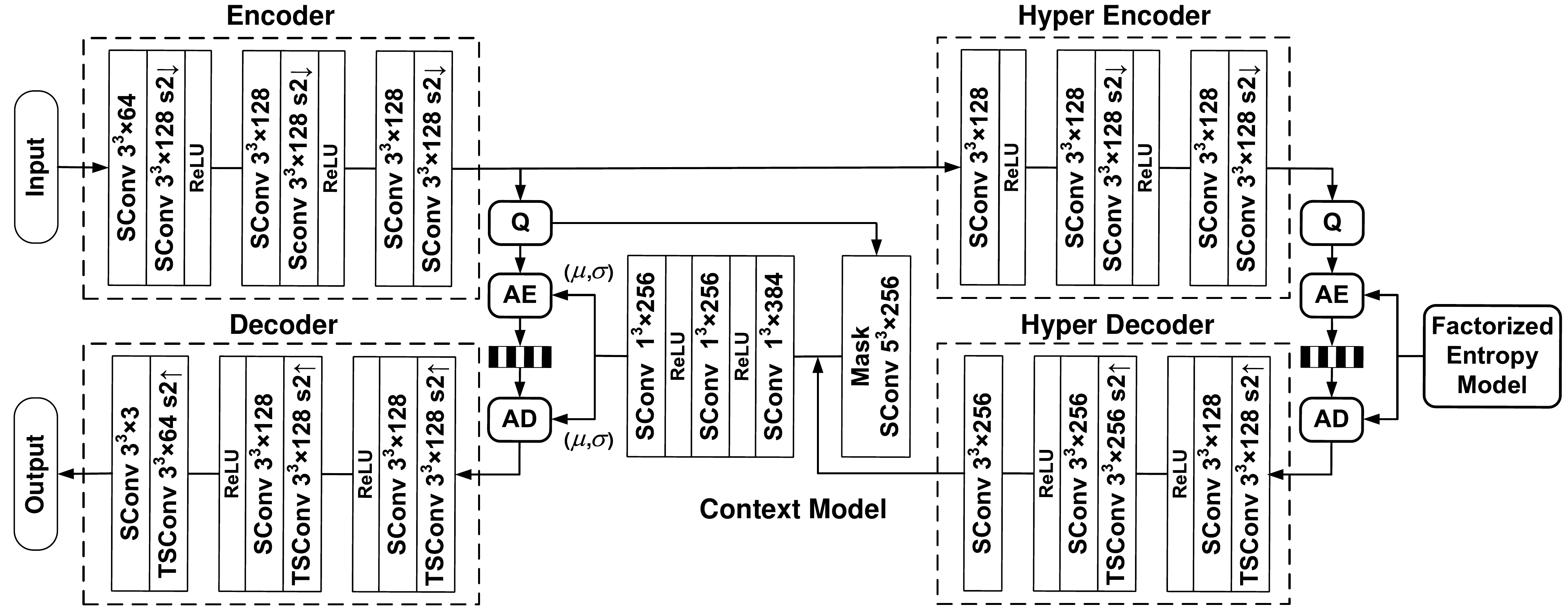}
	\end{center}
	\vspace{-0.1in}
	\caption{{\bf Sparse PCAC.} ``SConv ${n}^{3}\times {c}$'' and ``TSConv ${n}^{3}\times {c}$'' denote the sparse convolution and transposed sparse convolution with $c$ output channels and $n\times n\times n$ kernel size. ``$s\uparrow$'' and ``$s\downarrow$'' represent upscaling and downscaling at a factor of $s$ in each geometric dimension. ``ReLU'' stands for the Rectified Linear Unit. ``Q'' implies ``Quantization'', ``AE'' and ``AD'' are Arithmetic Encoder and Decoder, respectively.}
	\vspace{-0.1in}
	\label{fig:overview}
\end{figure*}

However,  very limited attention has been paid to learning-based attribute compression. Most of them are leveraging the popular models for geometry presentation to perform the attribute compression, such as the PointNet-style architecture in~\cite{Sheng2021DeepPCACAE}, 3D dense convolutions in~\cite{Alexiou2020TowardsNN}, implicit representation using  coordinate-based networks~\cite{Isik2021LVACLV}, {neural network based 3D to 2D projection in~\cite{Quach2020FoldingBasedCO}}, etc. 
Unfortunately, though these learnt PCAC solutions improve the efficiency of compression substantially, they are still inferior to the  conventional G-PCC with a noticeable performance gap.  Thus it is challenging to compactly represent 3D attributes due to the unconstrained placement of points in a free 3D space.

{\bf Our Approach.}
As motivated by the outstanding performance of the utilization of sparse tensor for representing the  point cloud geometry occupancy~\cite{Wang2020MultiscalePC,Wang2021SparseTM}, we extend it to characterize  the 3D color attributes of sparsely-distributed points, where a given point cloud can be represented using the sparse tensor $\{\vec{\bf C}, \vec{\bf F}\}$ with $\vec{\bf C} = \{(x_i,y_i, z_i)|~i\in[0,N-1]\}$ as the positively-occupied geometry coordinates associated with the corresponding color attributes $\vec{\bf F} = \{(R_i,G_i,B_i)|~i\in[0,N-1]\}$. $N$ is the total number of valid points. We assume losslessly-coded geometry and put the focus of this work on the coding of $\vec{\bf F}$ only.

To efficiently process the sparse tensor, we propose to apply the sparse convolutions as used in~\cite{Wang2020MultiscalePC,Wang2021SparseTM}. Specifically, we construct the VAE framework with paired encoder-decoder structure as in Fig.~\ref{fig:overview} to supervise the end-to-end (E2E) learnt PCAC. For convenience, we call it ``Sparse PCAC''.

The encoder applies stacked sparse convolution (and voxel downscaling) layers as the analysis transform to aggregate features as the compact representation of input 3D attributes at the bottleneck for compression, while the decoder mirrors the steps used in encoder as the synthesis transform for output reconstruction. To better compress the latent features, we use the hyper encoder-decoder to generate hyper priors that are coupled with the autoregressive neighbors to accurately estimate the conditional entropy probability. Such joint prior utilization shows convincing performance for the coding of 2D image/video~\cite{minnen2018joint,chen2021NLAIC}.

{\bf Contribution \& Prospect.}
%\begin{enumerate}
    1) This work is probably {the first attempt} to extend sparse convolutions for point cloud attribute compression. Together with the earlier works in~\cite{Wang2020MultiscalePC,Wang2021SparseTM}, we believe that the sparse tensor (with sparse convolutional computation) is a powerful tool for efficient point cloud representation and compression;  2) Even though we simply stack sparse convolutions (and resampling) for PCAC, the results show encouraging compression efficiency with {\it noticeable quantitative gains to standardized G-PCC version 6 and other learning-based method}, and {\it competitive qualitative evaluation with the latest G-PCC version 14}; 3) We are confident that the proposed Sparse PCAC will outperform the G-PCC version 14 if we further include the multiscale and multistage mechanism~\cite{Wang2021SparseTM} to massively exploit the cross-scale and cross-stage correlations for prediction. One evidence that the G-PCC version 14 greatly improves the G-PCC version 6 is the use of the transform coefficient prediction~\cite{gpcc2021description}.
%\end{enumerate}

\Section{Related Work} \label{sec:related}

We mainly review research activities on  point cloud attribute compression and sparse tensor based geometry compression that are closely related to this work.

\textbf{Point Cloud Attribute Compression (PCAC).}
Over the past years, many transformation methods have been proposed for PCAC. For instance, Graph Fourier Transform (GFT) is a popular tool for decorrelating unstructured signal, which was used to convert point cloud attribute from the spatial domain into the spectral domain for efficient compression~\cite{Zhang2014PointCA}. But, GFT is computationally expensive, thus Queiroz and Chou~\cite{Queiroz2016CompressionO3} proposed a computationally-efficient Region-Adaptive Hierarchical Transform (RAHT) that is later adopted into the MPEG G-PCC~\cite{gpcc2021description} as the core transformation tool for attribute compression (a.k.a., RAHT anchor in Sec.~\ref{sec:perf}). In principle, the RAHT is a hierarchical sub-band transform that resembles an adaptive variation of a Haar wavelet. Then, G-PCC was improved with better entropy coding\footnote{https://github.com/MPEGGroup/mpeg-pcc-tmc13} in version 6 (or TMC13v6 anchor as of July 2019) and further with the prediction of RAHT coefficients in the latest version 14 (or TMC13v14 anchor as of October 2021) that reports the state-of-the-art PCAC efficiency.
% \footnote{We interchangeably use G-PCC version 6 or TMC13 v6 (G-PCC version 14 or TMC13 v14).}

Besides these rules-based approaches, few learning-based PCAC solutions have emerged very recently.
For example,  Sheng~\textit{et al.}~\cite{Sheng2021DeepPCACAE} proposed a PointNet style autoencoder for attribute compression (e.g., Deep-PCAC in Sec.~\ref{sec:perf}), and Alexiou~\textit{et al.}~\cite{Alexiou2020TowardsNN} applied 3D dense convolutional autoencoder for the coding of both geometry and attribute components. However, they both suffer inferior performance with a significant loss to the G-PCC.

{Different from the aforementioned end-to-end learnt PCAC, Quach~\textit{et al.}~\cite{Quach2020FoldingBasedCO} proposed to learn a neural network(NN)-based mapping function to convert a 3D point cloud into a 2D grid, and then apply conventional image codec to encode projected attribute. Such 3D-to-2D projection based PCC exploration follows the direction of video-based PCC~\cite{graziosi2020overview}.
Recently, Isik~\textit{et al.}~\cite{Isik2021LVACLV} proposed to use a coordinate-based/implicit network for attribute compression with the coefficients of RAHT as an additional latent vector. These methods offer alternative solutions for learnt PCAC. But the coding efficiency still suffers and the model generalization is challenging, e.g., Quach~\textit{et al.}~\cite{Quach2020FoldingBasedCO} had to overfit the mapping function for individual sample, and Isik~\textit{et al.}~\cite{Isik2021LVACLV} had to re-optimize the network through back-propagation for each point cloud. The poor model generalization hinders the application in practice.}

\textbf{Sparse Tensor-based PCGC.}
The sparse tensor is used for efficient convolution on spatially sparse data like voxelized point clouds~\cite{Choy20194DSC}. Different from the volumetric model which stores all voxels in a 3D volume, sparse tensor only stores the occupied voxels, which can be organized using a set of coordinates $\vec{\bf C}$ and associated attributes $\vec{\bf F}$. 
The sparse tensor uses a hash map and auxiliary table to index valid voxels and manage their connections efficiently, which can be used to construct network layers using convolution, pooling, etc, as the use of normal dense tensor, and greatly reduce the computational and space complexity to the regular dense convolution neural networks because of limited computation on sparsely occupied voxels. The representative sparse tensor  library is the MinkowskiEngine\footnote{https://github.com/NVIDIA/MinkowskiEngine}.

Wang~\textit{et al.}~\cite{Wang2021SparseTM,Wang2020MultiscalePC} have introduced the sparse tensor representation for point cloud geometry compression (PCGC) where sparse convolutions are used to perform the cross-scale and cross-stage prediction to effectively exploit the redundancy. Such Sparse PCGC reported the state-of-the-art performance~\cite{Wang2021SparseTM}, motivating us to extend the sparse tensor for the compression of attribute information.

\section{Sparse PCAC}
This section first details the general VAE framework for E2E Sparse PCAC as in Fig~\ref{fig:overview}, and then describes the context modeling with joint autoregressive and hyper priors.

\subsection{Sparse Tensor-based VAE Architecture}
The proposed method reuses the effective  and prevailing VAE structure~\cite{minnen2018joint, chen2021NLAIC}, and applies the sparse tensor and sparse convolution for data representation and processing. As aforementioned, for a given point cloud, we have the full knowledge of its  geometry coordinates $\vec{\bf C}$ of occupied voxels. This work only considers the compression of attribute information $\vec{\bf F}$. 
% Following the conventions, we typically voxelize the input $\vec{\bf F}$ using its volumetric representation \textbf{X}. 
Following the conventions, we use \textbf{X} and \textbf{\^{X}} to denote the input attribute and the reconstructed output attribute of the sparse tensor. Both of them are vectorized triple-channel (e.g., RGB) color attributes.
% training dataset
\begin{figure}[t]
\centering
\subfloat[]{\includegraphics[width=3in]{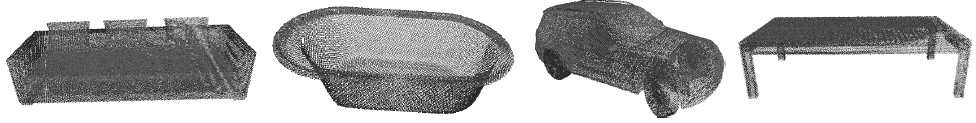}\label{fig:shapenet}}\\\vspace{-0.04in}
\subfloat[]{\includegraphics[width=3in]{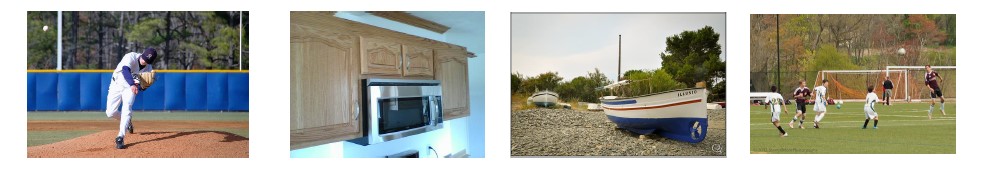}\label{fig:coco}}\\\vspace{-0.04in}
\subfloat[]{\includegraphics[width=3in]{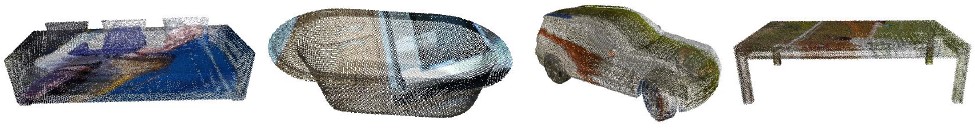}\label{fig:synthetic}}\vspace{-0.04in}
\caption{{\bf Training Dataset.} (a) ShapeNet~\cite{Chang2015ShapeNetAI}. (b) COCO~\cite{Lin2014MicrosoftCC}. (c) {Synthetic colored point clouds for training.}}
\vspace{-0.1in}
\label{fig:training-dataset}
\end{figure}
% figs: testing dataset
\begin{figure}[t]
\centering
\includegraphics[width=3in]{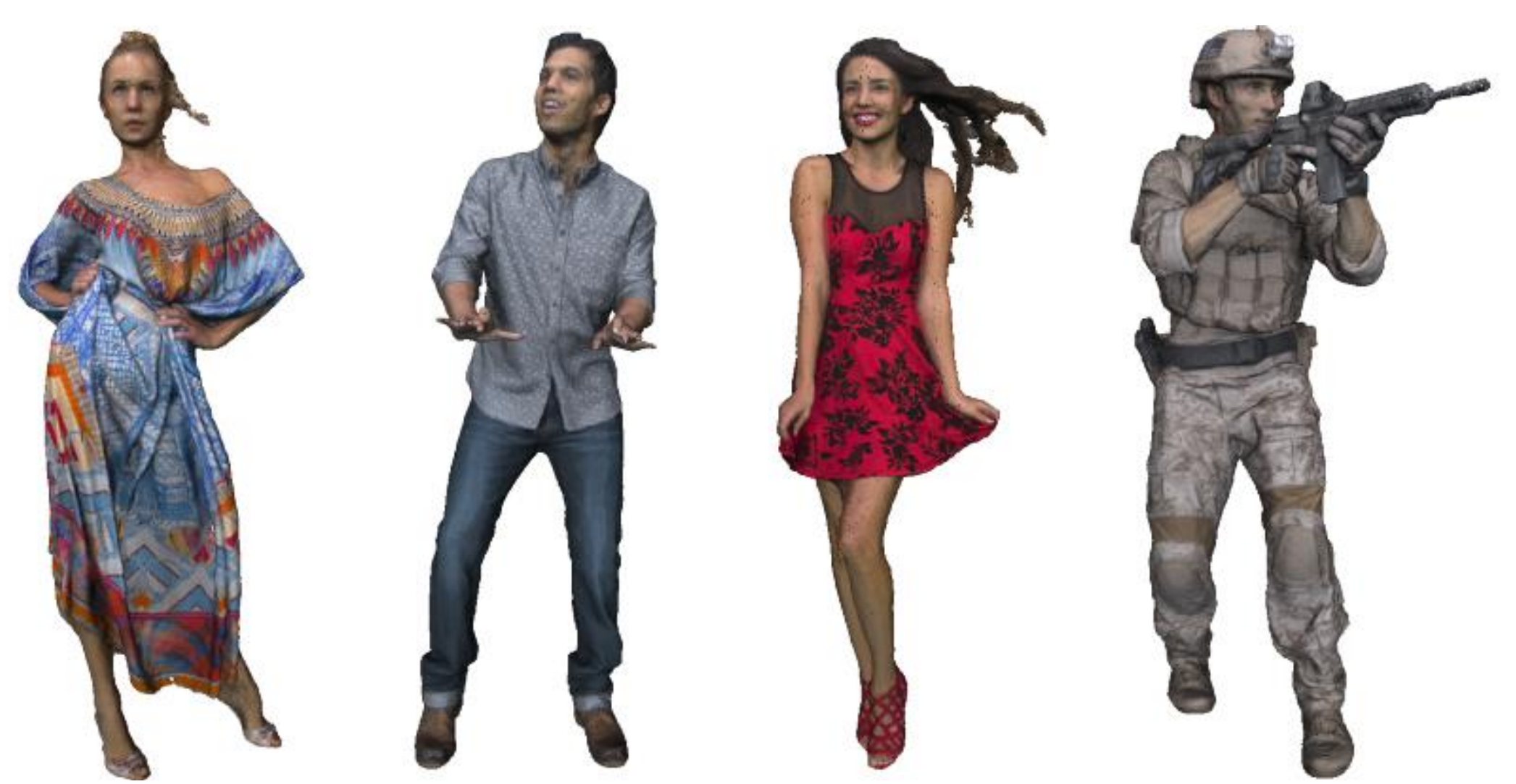}
\caption{{\bf Testing Dataset.} longdress, loot, redandblack, soldier.}
\vspace{-0.1in}
\label{fig:testing-dataset}
\end{figure}

The general framework contains two parts. The first is the main \textit{Encoder} and \textit{Decoder} for analysis and synthesis transform. It learns a quantized latent representation of original input, i.e., \textbf{\^Y}. The second part is responsible for devising a probabilistic model over the quantized latents for entropy coding. It combines the autoregressive neighbors and hyper priors generated by a pair of \textit{Hyper Encoder} and \textit{Hyper Decoder} to derive the mean and scale parameters $(\mu, \sigma)$ assuming the conditional Laplacian distribution. Then, \textbf{\^Y} is encoded into the bitstream with estimated probability through the arithmetic coding engine. 

%We adopt sparse tensor networks for efficient and low-complexity processing. Details about the individual layers in each module are presented in Fig.~\ref{fig:overview}. 
The encoder consists of 6 Sparse Convolutional (SConv) layers and 2 ReLU activation layers. All SConv layers have the kernel size of $3\times3\times3$ for neighboring features aggregation, and most of them have 128 channels for high-dimension feature embedding. Three SConv layers have a stride of two for down-scaling at each geometric dimension. At the output of encoder, latent features \textbf{\^Y} is at a geometric size of $\frac{1}{8}\times\frac{1}{8}\times\frac{1}{8}$ of the original \textbf{X}, with 128 channels.
%latent features with a size of $\frac{1}{8}\times\frac{1}{8}\times\frac{1}{8}$ of the original geometric scale and 128 channels.
The decoder has a mirroring network structure of encoder to generate RGB color reconstruction \textbf{\^{X}} from latents \textbf{\^Y}. It uses Transposed Sparse Convolution (TSConv) with a stride of two for up-scaling (e.g., $\times2$ in each geometric dimension).

The hyper encoder/decoder pair learns \textit{Hyper priors} from latents \textbf{\^Y} to improve the context modeling. It shares a similar structure to the main encoder/decoder network but only has two down-/up-scaling layers. 
The context model uses a $5\times5\times5$ Masked Sparse Convolution (Masked SConv) layer to model probability using causal neighbors in a $5\times5\times5$ receptive field. The output of Masked SConv is combined with the hyper prior by stacked $1\times1\times1$ SConv layers to generate the mean and scale parameters for probability estimation used in arithmetic encoding and decoding.

\subsection{Entropy Model}
Accurate context modeling is critical for entropy coding.
Numerous works have been developed, e.g., factorized entropy model~\cite{balle2018variational}, conditional entropy models based on hyperprior, autoregressive prior or their combination~\cite{minnen2018joint,chen2021NLAIC},  and successfully applied to well structured 2D image data. It still lacks exploration on unstructured, sparsely-scattered 3D point cloud attribute. We jointly utilize the autoregressive neighbors and hyper priors in  this work to exploit the correlation for the efficient compression of latents \textbf{\^Y}.

\textbf{Autoregressive Priors.}
Local neighboring voxels usually present high correlations. Inspired by Masked CNN based spatial context model in image compression~\cite{minnen2018joint,chen2021NLAIC}, we use a $5\times5\times5$ Masked SConv to exploit the spatial correlation on \textbf{\^Y}, where the probability distribution of current element $\hat{y}_{i}$ can be modeled on the causal elements in an autoregressive way.

\textbf{Hyper Priors.}
We also use a hyper encoder/decoder network to characterize the hyper priors \textbf{\^Z} that are compressed with a \textit{factorized entropy model}~\cite{balle2018variational}, as the conditional information to help the entropy modeling of \textbf{\^Y}.

\textbf{Joint Autoregressive \& Hyper Priors.}~\label{subsec:entropy-model}
Autoregressive and hyper priors are complementary and can be combined to jointly exploit the probabilistic structure in the latents better than using one of them alone. We combine their information in the context model $g_{cm}$ using stack $1\times1\times1$ SConvs to generate the mean and scale parameters $(\mu, \sigma)$ for a conditional Laplacian entropy model. 
Thus the probability $p_{\hat{y}}$ of the elements $\hat{y}$ in latents are conditioned on  $\hat{z}$ and causal neighbors, and  $p_{\hat{y}}$ can be estimated by cumulation over the Laplacian distribution $\mathcal{L}$ as shown in Eq.~\ref{eq:entropy-model}, 
\begin{align}
    \vspace{-0.05in}
    p_{\hat{y}|\hat{z}}(\hat{y}|\hat{z},\theta_{hd}, \theta_{cm}) 
    =\prod_{i}\Bigl(\mathcal{L}(\mu_{i}, \sigma_{i})*\mathcal{U}(-\tfrac{1}{2},\tfrac{1}{2})\Bigr)(\hat{y}_{i})\nonumber\\
    \text{with~}\mu_{i}, \sigma_{i}=g_{cm}(\hat{y}<i,\psi_{i};\theta_{cm}), 
    \psi_{i}=g_{hd}(\hat{z};\theta_{hd})
\label{eq:entropy-model}
\vspace{-0.1in}
\end{align} 
where $\mathcal{U}(-\tfrac{1}{2},\tfrac{1}{2})$ represents uniform distribution ranging from $-\tfrac{1}{2}$ to $\tfrac{1}{2}$, 
$g_{cm}$ and $g_{hd}$ are the context model and hyper decoder with the parameters $\theta_{hd}$ and $\theta_{cm}$, $\hat{y}<i$ and $\hat{z}$ represent the causal elements and hyperprior respectively.
% \begin{equation}
%     p_{\hat{z}|\phi}(\hat{z}|\phi) =\prod_{i} \Bigl(p_{\hat{z}_{i}|\phi^{(i)}}(\phi^{(i)}) * \mathcal{U}( -\tfrac{1}{2},\tfrac{1}{2})\Bigr)(\hat{z}_{i}),
% \end{equation}

% figs: RD-curves
\begin{figure}[t]
\centering
\includegraphics[width=1.6in]{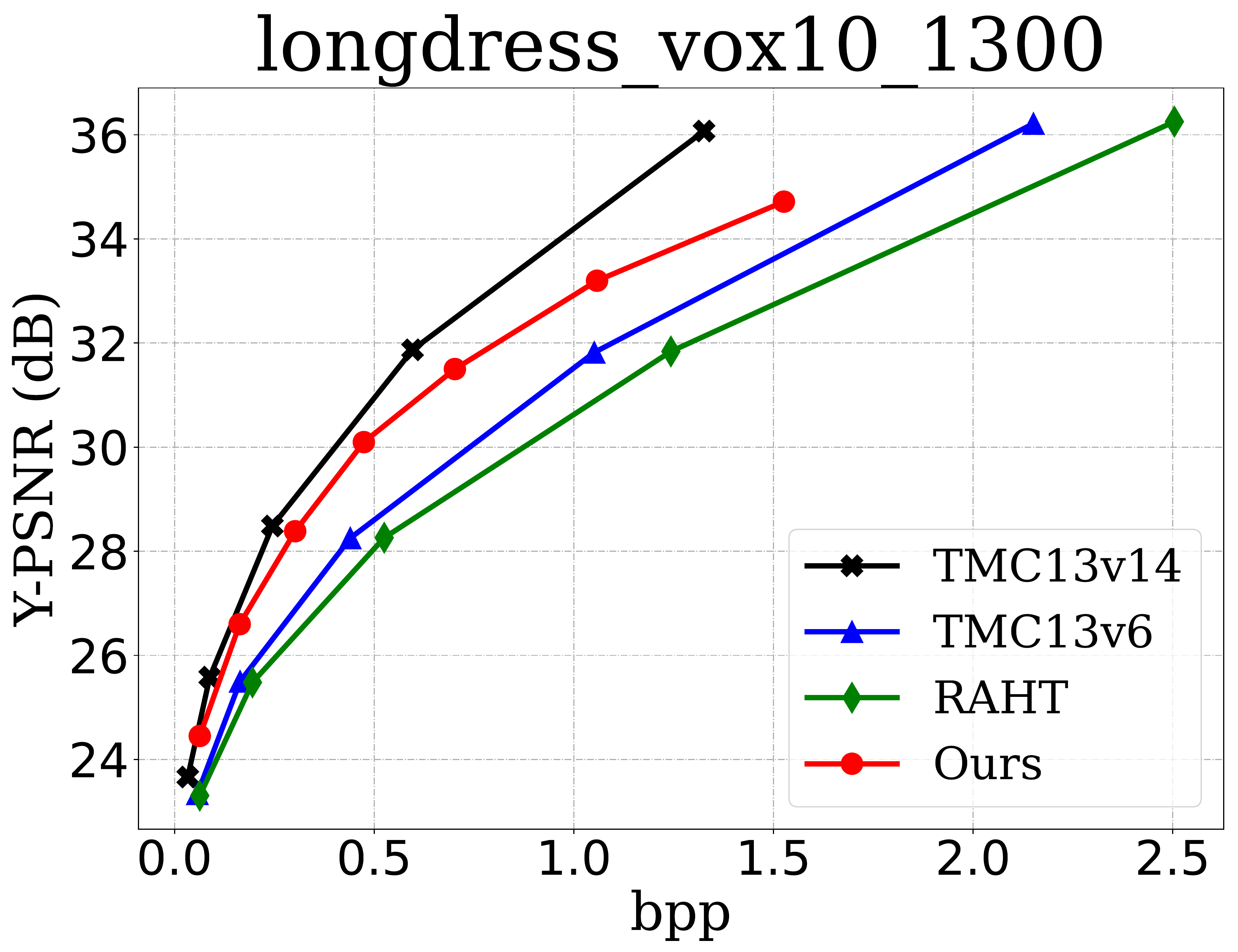}
\includegraphics[width=1.6in]{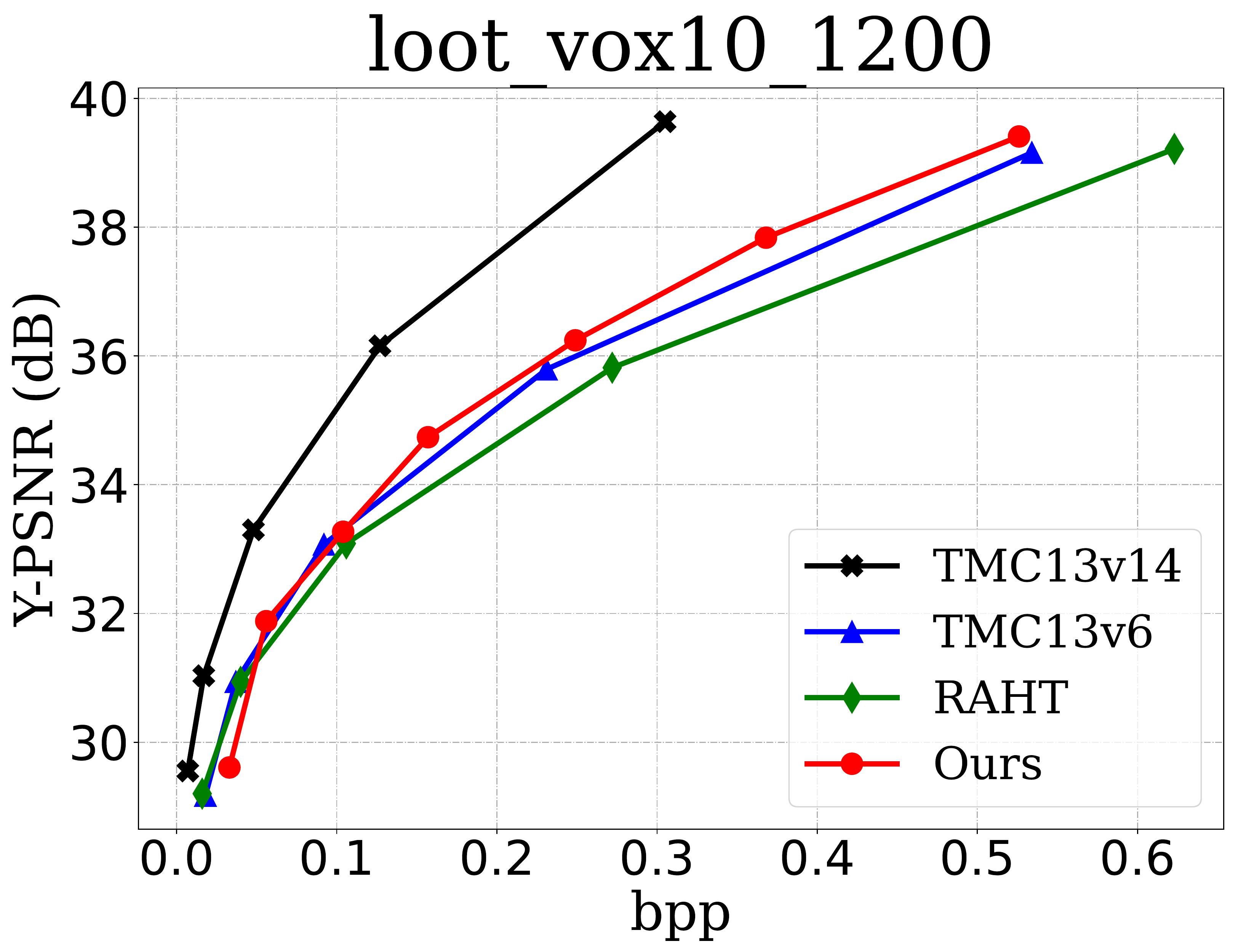}
\includegraphics[width=1.6in]{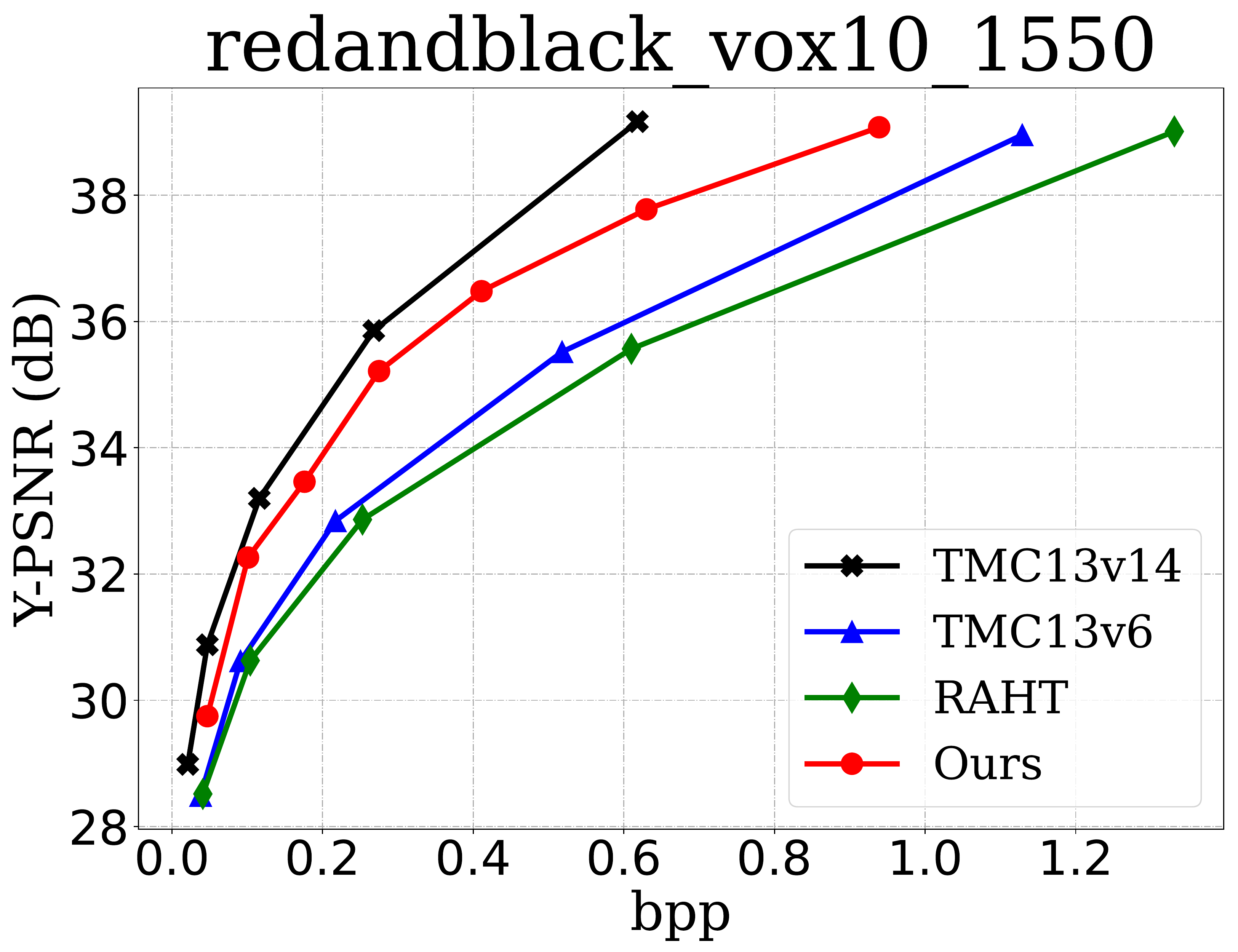}
\includegraphics[width=1.6in]{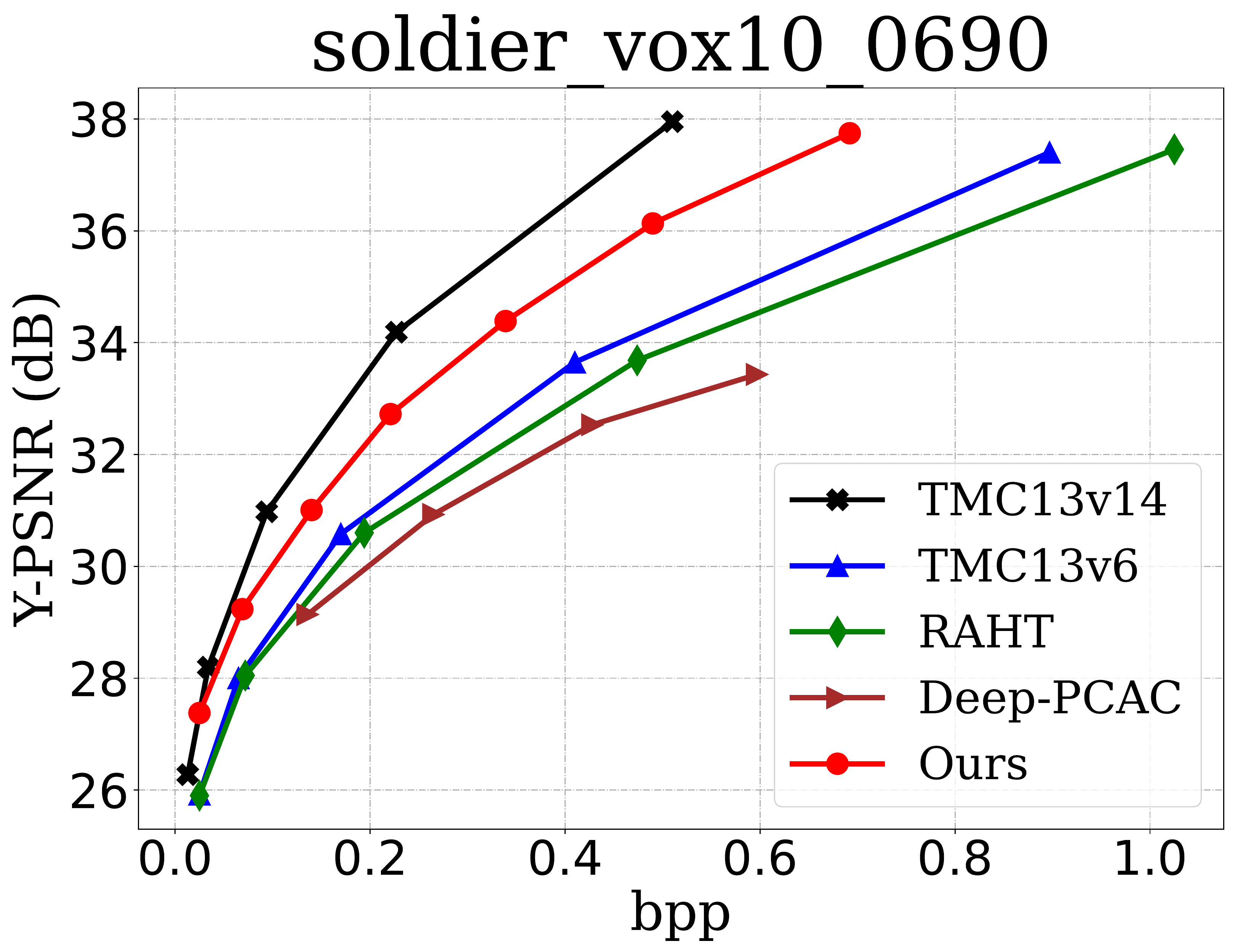}
\caption{{Performance comparison using rate-distortion curves.}}
\vspace{-0.1in}
\label{fig:rdcurves}
\end{figure}

% Table: BD-Rate
% \begin{table}[t]\footnotesize
% \caption{{BD-BR(\%) and BD-PSNR(dB) comparisons of the proposed method with TMC13 and RAHT on Y channel.}}
% \label{table:BDBR}
% \centering
% \begin{tabular}{|c| c | c | c |}
% \hline
% {\textbf{PCs}} & {\textbf{TMC13v14}} & {\textbf{TMC13v6}} &  {\textbf{RAHT}} \\
% % \cline{2-4}
% % & BD-BR/BD-PSNR & BD-BR/BD-PSNR & BD-BR/BD-PSNR \\ 
% \hline
% longdress & 38.7\%/-1.03dB & -25.1\%/+0.94dB & -36.5\%/+1.51dB\\
% loot & 119.8\%/-2.31dB & ~3.7\%/-0.16dB & ~-7.3\%/+0.13dB \\
% red\&black  &64.2\%/-1.49dB & -21.5\%/+0.74dB & -31.6\%/+1.16dB  \\
% soldier & 39.7\%/-1.03dB & -32.4\%/+1.22dB & -39.8\%/+1.54dB \\
% \hline
% \textbf{Average} & \textbf{65.6\%}/\textbf{-1.46dB} & \textbf{-18.8\%}/\textbf{+0.68dB}  & \textbf{-28.8\%}/\textbf{+1.08dB} \\
% \hline
% \end{tabular}
% \end{table}

\begin{table*}[t]\footnotesize
\caption{{BD-BR(\%) and BD-PSNR(dB) comparisons of the proposed method with TMC13 and RAHT.}}
\label{table:BDBR}
\centering
\begin{tabular}{|c| c| c| c| c | c| c| c| c | c| c| c| c|}
\hline
\multirow{3}{*}{Point Cloud} & \multicolumn{4}{|c|}{\textbf{TMC13v14}} & \multicolumn{4}{|c|}{\textbf{TMC13v6}} & \multicolumn{4}{|c|}{\textbf{RAHT}} \\
\cline{2-13}
& \multicolumn{2}{|c|}{BD-BR (\%)} & \multicolumn{2}{|c|}{BD-PSNR (dB)} & \multicolumn{2}{|c|}{BD-BR (\%)} & \multicolumn{2}{|c|}{BD-PSNR(dB)} & \multicolumn{2}{|c|}{BD-BR (\%)} & \multicolumn{2}{|c|}{BD-PSNR (dB)} \\ 
\cline{2-13}
& Y & YUV & Y & YUV & Y & YUV & Y & YUV & Y & YUV & Y & YUV \\ 
\hline
longdress & +27 & +27 & -0.86 & -0.81 & -31 & -32 & +1.23 & +1.22 & -42 & -42 & +1.76 & +1.72 \\
loot & +118 & +142 & -2.20 & -2.50 & +2 & +17 & +0.00 & -0.40 & -9 & +6 & +0.40 & -0.10 \\
red\&black & +35 & +55 & -0.89 & -1.28 &  -36 & -26 & +1.35 & +0.90 & -44 & -36 & +1.76 & +1.30 \\
soldier & 38 &	66 & -1.05 & -1.60 & -33 & -17 & 1.28 & +0.50 & -41 & -25 & +1.59 & +0.75  \\
\hline
\textbf{Average} & +55\% & +72\% & -1.25dB & -1.55dB & \textbf{-24\%} & \textbf{-15\%} & \textbf{+0.97dB} & \textbf{+0.55dB} & \textbf{-34\%} & \textbf{-24\%} & \textbf{+1.38dB} & \textbf{+0.92dB} \\
\hline
\end{tabular}
\end{table*}

% figure: visualization
\begin{figure*}[t]
\centering
\includegraphics[width=6.5in]{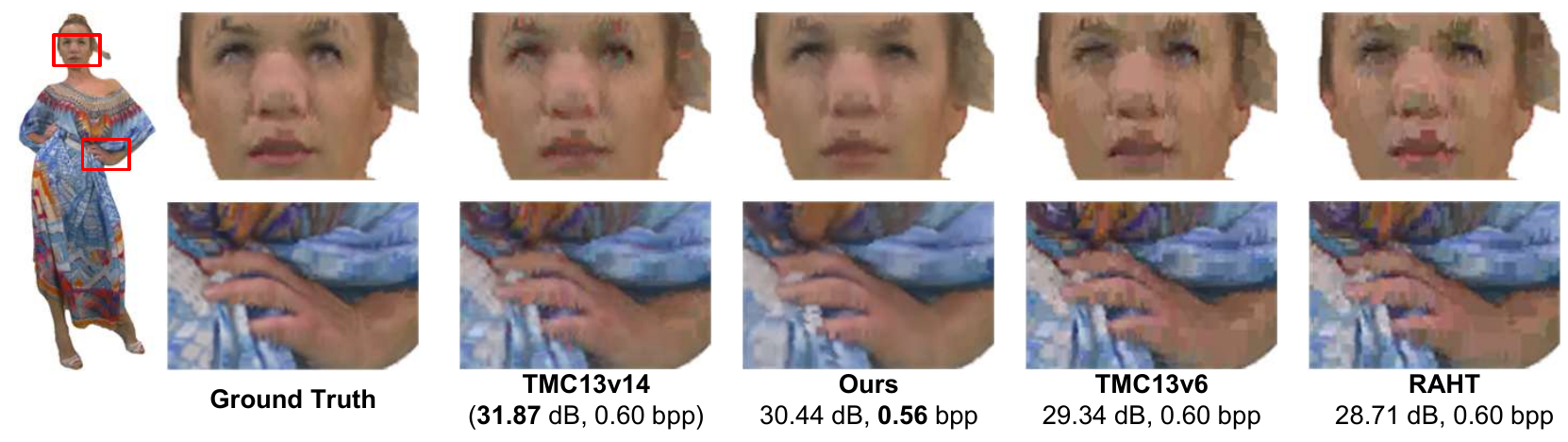}
\vspace{-0.1in}
\caption{{Qualitative visualization of reconstructed  ``longdress'' for ground
truth, TMC13v14, Ours, TMC13v6, and RAHT.}}
\label{fig:visualization}
\vspace{-0.1in}
\end{figure*}

\section{Experimental Results}
\subsection{Training}
\textbf{Dataset.} We synthesize the training dataset using popular ShapeNet~\cite{Chang2015ShapeNetAI} and COCO~\cite{Lin2014MicrosoftCC}, where ShapeNet provides geometry information, and COCO provides the attribute information. We first densely sample points on raw ShapeNet meshes, and randomly rotated and quantized the coordinates to 8-bit integers as in ~\cite{Wang2021SparseTM}. Next, we randomly select images from COCO and map them to the points as color attributes through projection. Fig.~\ref{fig:training-dataset} shows some examples of the training dataset and their corresponding geometry and attribute sources. We generated 6000 samples for training. 

\textbf{Loss Function.}
We use the Lagrangian loss for Rate-Distortion (R-D) optimization in training, 
% eq: loss function.
\begin{align}
    R + \lambda \cdot D =  R_{\hat{y}} +  R_{\hat{z}} + \lambda \cdot \sum_{j}\Vert x_{j} - \hat{x}_{j} \Vert_{2}^{2}. \label{eq:rd-loss}
    \end{align}
where $\lambda$ is the factor that determines the R-D trade-off. $R_{\hat{y}}$ = $\sum_{i}-\log_{2}(p_{\hat{y}_{i}|\hat{z}_{i}}(\hat{y}_{i}|\hat{z}_{i}))$ is the bitrate of latents \textbf{\^Y}, while $R_{\hat{z}}$  = $\sum_{i}-\log_{2}(p_{\hat{z}_{i}|\phi^{(i)}}(\hat{z}_{i}|\phi^{(i)}))$ is the bitrate of hyperprior \textbf{\^Z}. %, which are approximated by the cross entropy loss in Eq.~\ref{eq:latent-rate}  and Eq.~\ref{eq:prior-rate}.
$p_{\hat{y}_{i}|\hat{z}_{i}}$ is the conditional probability model of \textbf{\^Y} in \eqref{eq:entropy-model}, and $p_{\hat{z}_{i}|\phi^{(i)}}$ is the factorized probability model~\cite{balle2018variational} of \textbf{\^Z}. The distortion is measured by the Mean Squared Error (MSE) between original input \textbf{X} and reconstructed output \textbf{\^X} in the YUV colour space.

\textbf{Training Procedure.} We set $\lambda$ to 16000, 8000, 4000, 2000, 1000, 400, 100 to obtain models with different bitrates. Each model is trained on the dataset for 50 epochs, with the learning rate decayed from 1e-4 to 2e-5. Due to the lightweight model structure and efficient training settings, it only takes several hours to train each model on one Nvidia GeForce RTX 3090 GPU.

% RD-curves
\begin{figure}[t]
\centering
\includegraphics[width=2.6in]{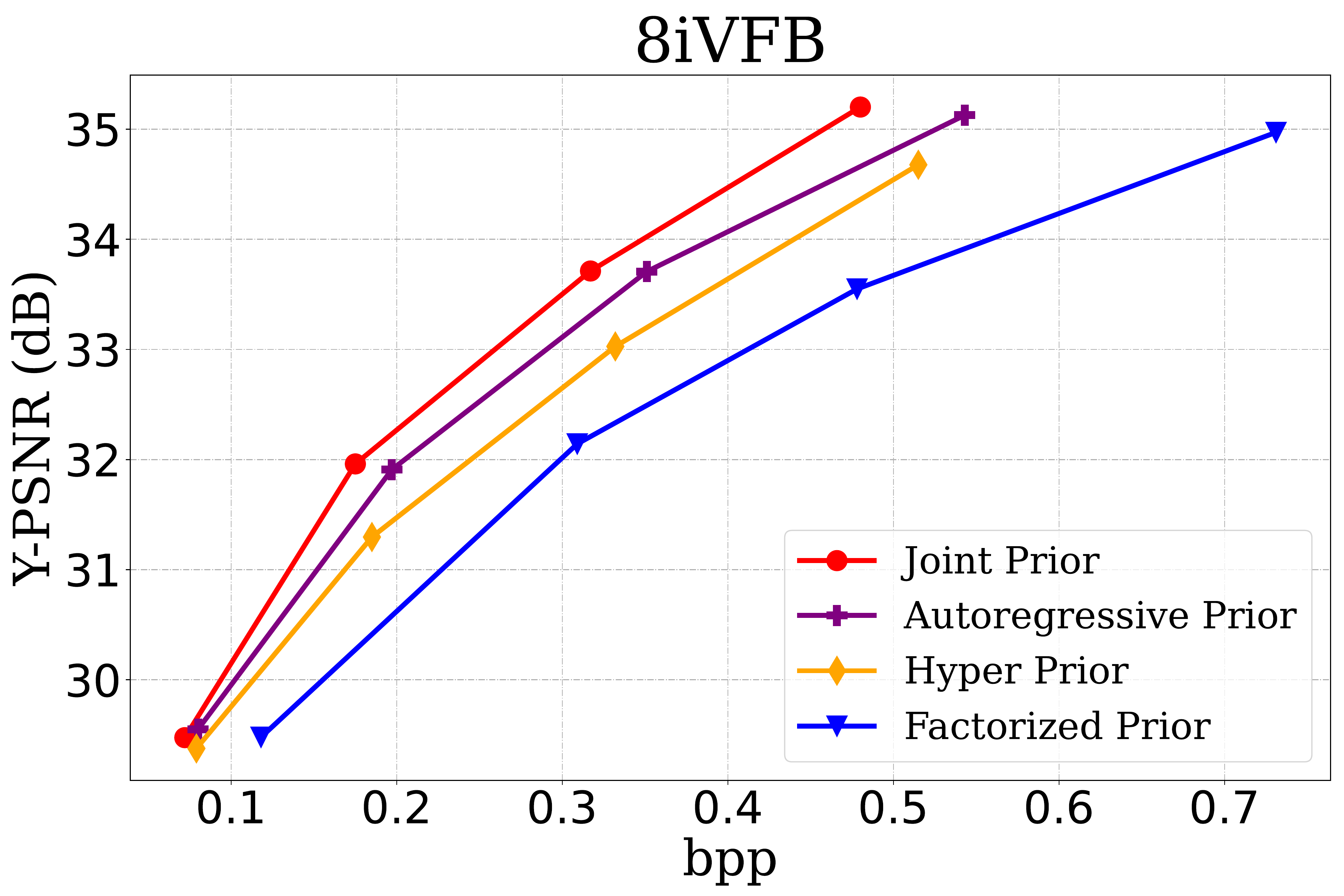}
\vspace{-0.1in}
\caption{{Impact of priors for entropy modeling.}}
\label{fig:ablation}
\vspace{-0.1in}
\end{figure}
\SubSection{Performance Evaluation} \label{sec:perf}
\textbf{Dataset \& Anchors}
We evaluate the proposed method using 8i Voxelized Full Bodies (8iVFB)~\cite{8i20178i} that are chosen as the standard test set in MPEG, JPEG, etc. As in Fig.~\ref{fig:testing-dataset}, they have quite different geometry and attribute characteristics from the training dataset, which can help to measure the model generalization.  We compare with three different G-PCC versions, including its original RAHT implementation\footnote{https://github.com/digitalivp/RAHT}, the TMC13v6 with better entropy coding, and the latest and state-of-the-art TMC13v14 with RAHT coefficients prediction on top of the TMC13v6. We have followed the common test conditions~\cite{MPEG_GPCC_CTC} to generate the anchor results.

\textbf{Quantitative Comparison.}
We follow the common practice to measure the rate dissipated for all color channels of each point (i.e., bits per points, bpp) and distortion in terms of PSNR in dB of the Y and YUV channnels, which are computed using MPEG PCC pc\_error tools. We use Bjøntegaard Delta PSNR (BD-PSNR) (in dB) and Bjøntegaard Delta bit rate (BD-BR) (in percentage) to measure averaged R-D performance.

As shown by the R-D curves in Fig.~\ref{fig:rdcurves} and the BD-BR/BD-PSNR gains in Tab.~\ref{table:BDBR}, we outperform TMC13v6 and RAHT largely, having 24\% BD-BR reduction and 0.97dB BD-PSNR improvement than TMC13v6, 34\% BD-BR reduction and 1.38dB BD-PSNR gain over RAHT, measured on Y channel. But we currently have performance loss compared to the TMC13v14.
Besides, we compare our method with a learning-based method Deep-PCAC~\cite{Sheng2021DeepPCACAE} on the common test sample ``soldier'' since the other three point clouds are used for training by Deep-PCAC. A significant gain of our method over Deep-PCAC is clearly observed. 
% Another E2E learned PCAC method~\cite{Alexiou2020TowardsNN} based on 3D CNN is inferior to RAHT. 
Due to the generalization problem of the methods of Quach~\textit{et al.}~\cite{Quach2020FoldingBasedCO} and Isik~\textit{et al.}~\cite{Isik2021LVACLV}, the comparison would be unfair, so we do not compare with them here.

\textbf{Qualitative Evaluation.} 
We visualize the reconstructed color of different PCAC methods in Fig.~\ref{fig:visualization}.
Our method provides the closest appearance to the ground truth with more smooth textures incurred by the MSE loss. Noticeable blocking artifacts and noise are seen in reconstructed snapshots of RAHT, TMC13v6, and even TMC13v14. As seen, our method offers competitive visual presentation to the TMC13v14.
% and maybe it can be further improved by using better distortion metrics that fit the human visual system as the loss function.
%, and RAHT, G-PCCv6, and even G-PCCv14 have significant noise and blocking artifacts, while ours show more blurring artifacts and smoother texture, this may be due to the influence of MSE loss and can be improved by using other loss functions that conform human visual system.

{\bf Ablation Studies.} We examine the efficiency of the entropy model with the utilization of different priors. We take the simple factorized entropy model without any context as the baseline. As shown in Fig.~\ref{fig:ablation}, the context model based on autoregressive prior and hyper prior both yield significant gains over the baseline, i.e, 31.2\% and 21.5\% gains. When we combine them together, the joint prior can achieve 38.9\% gains. This fully demonstrates the effect of priors on improving the context modeling for entropy coding.

\section{Conclusion}
This work has demonstrated the powerful capacity of sparse tensor representation for point cloud attribute compression. We simply stack sparse convolutional layers in a VAE framework for attribute coding where joint autoregressive and hyper priors are utilized for entropy modeling of the latents. The proposed method outperforms the TMC13v6 and RAHT with 24\% and 34\% BD-Rate improvements on Y channel and also shows competitive qualitative visualization to the state-of-the-art TMC13v14. 

As seen, the introduction of coefficients prediction in TMC13v14 greatly improves the TMC13v6. We expect the significant performance improvement of proposed Sparse PCAC if we further include the cross-scale and cross-stage prediction as studied in~\cite{Wang2021SparseTM} to massively exploit the redundancy.
\bibliographystyle{latex8}
%\bibliography{0_main}
\bibliography{0_main.bbl}
\end{document}